\documentclass[twocolumn,showpacs,preprintnumbers,amsmath,amssymb]{revtex4}

\usepackage{graphicx}
\usepackage{amsfonts,latexsym}
\begin{document}

\title{
Ground state entanglement of the BCS model
} 

\author{Clare Dunning}
%\email{tcd@maths.uq.edu.au}
\author{Jon Links} 
%\email{jrl@maths.uq.edu.au} 
\author{Huan-Qiang Zhou} 
%\email{hqz@maths.uq.edu.au}
\affiliation{Centre for Mathematical Physics, School of Physical
Sciences, The University of
Queensland, Brisbane, 4072, Australia}

\begin{abstract}
The concept of local concurrence is used to quantify the entanglement
between a single qubit and the remainder of a multi-qubit system. 
For the ground state of the BCS model in the thermodynamic limit 
the set of local concurrences completely describe the entanglement.  
As a measure for the entanglement of the full system we investigate 
the Average Local Concurrence (ALC). 
We find that the ALC  
satisfies a simple relation with 
the order parameter.
We then show that for  
finite systems with fixed particle
number, a relation between the ALC and the condensation energy 
exposes a threshold coupling. Below the threshold,  
entanglement measures besides the ALC 
are significant.

\end{abstract}
\pacs{03.65.Ud, 74.20.Fg} 

\maketitle  
                     
%************************** Text Begins here ******************************

%  Greek letters
\def\aa{\alpha} 
\def\bb{\beta}
\def\a{\hat a}
\def\b{\hat b}
\def\d{\dagger}
\def\de{\delta} 
\def\e{\epsilon}
\def\ve{\varepsilon}
\def\g{\gamma}
\def\K{\kappa}
\def\ap{\approx}
\def\l{\lambda}
\def\o{\omega}
\def\t{\tilde{\tau}}
\def\s{\sigma}
\def\D{\tilde{\Delta}}
\def\L{\Lambda}
\def\T{{\cal T}}
\def\TT{{\tilde{\cal T}}}
\def\E{{\cal E}} 
\def\f{\overline{f}}
\def\q{\overline{q}}
\def\tp{\otimes}
\def\I{{\rm id}}
\def\rar{\rightarrow}
\def\C{\tilde{C}}
% Shorthands for \begin{equation} and the like

\def\beq{\begin{equation}}
\def\eeq{\end{equation}}
\def\bea{\begin{eqnarray}}
\def\eea{\end{eqnarray}}
\def\ba{\begin{array}}
\def\ea{\end{array}}
\def\no{\nonumber}
\def\le{\langle}
\def\re{\rangle}
\def\lt{\left}
\def\rt{\right}
\def\o{\omega}
\def\d{\dagger}
\def\nn{\nonumber}
\def\j{{ {\cal J}}}
\def\n{{\hat n}}
\def\N{{\hat N}}
\def\A{{\cal A}}
\def\TT{{\tilde {\cal T}}}
\def\bg{\tilde{\Delta}}
\def\CC{\mathcal{C}}
\def\oC{\overline{C}} 
\def\E{\mathbb{E}} 
\def\EE{\mathcal{E}}
\def\tE{\tilde{E}}
\newcommand{\reff}[1]{eq.~(\ref{#1})}

Quantification of entanglement remains as a major challenge in quantum
information theory. 
For a state of a bi-partite system, it is well
known that a measure for the entanglement between the two subsystems is
given by the von Neumann entropy \cite{pr}. 
For a two-qubit system, the
concurrence \cite{hw} has been introduced as an alternative measure which
is related to the von Neumann entropy in a bijective manner. 
In the
language of spin-1/2 particles, concurrence can be described in terms of a
time-reversal operation. 
Using this concept, generalisations of
concurrence have been proposed for multi-qubit systems and in particular
they have been applied to quantify the entanglement of the ground state of
the BCS model in the thermodynamic limit \cite{g,md}, which breaks
time-reversal symmetry due to broken gauge symmetry.

In this Letter we adopt a different approach to investigate the ground
state entanglement of the BCS model. 
We will use the notion of \emph{local concurrence} which is defined 
in analogy with the functional relation that
exists between concurrence and von Neumann entropy for a two qubit system
(cf. \cite{ckw}).
It is a measure of the entanglement between a single
qubit and the remainder of the system. 
We then define an 
entanglement measure which is the Average Local Concurrence (ALC). 
The ALC
satisfies the properties of an entanglement monotone (EM): 
it vanishes for a state if and only if that state is a product
state, is invariant under local unitary transformations, and does not
increase on average under local operations assisted by classical
communication.  

For multi-qubit systems the number of independent EMs 
is the same as the number of 
non-local invariants \cite{ging}, growing
exponentially with system size ($2^{L+1}-3L-2$ 
where $L$ is the number of qubits
\cite{lps,kem}). 
It is therefore useful to identify
EMs which can be related to physical aspects of the system. 
In the thermodynamic   
limit we will show that the ALC 
for the ground state of the BCS model displays 
a simple relationship with the magnitude of the order parameter. 
For finite systems the order parameter vanishes, 
so we may use the ALC in lieu of the order parameter. 
Our investigation of a relationship between the ground state ALC and the
condensation energy exposes a threshold coupling  
which signifies the onset of entanglement not measured by the local
concurrences. 

To study the entanglement of any pure state we can examine the von Neumann
entropy. 
Consider a general quantum system comprised of $L$ qubits and a
partition into two subsystems denoted $A$ and $B$. 
For any pure state
density matrix $\rho$ the entanglement
$\EE_{AB}(\rho)$ between $A$ and $B$ is given by the von Neumann
entropy
$$\EE_{AB}(\rho)=-{\rm tr}(\rho_A \log \rho_A)=-{\rm tr}(\rho_B \log
\rho_B) $$
where the logarithm is taken base 2 and $\rho_A$ is the reduced density
matrix obtained from $\rho$ by taking the partial trace over the state
space of subsystem $B$. 
The reduced density matrix $\rho_B$ is defined
analogously. 
 
Now we make precise the definition of local concurrence in a general
context. 
Hereafter we will only deal with the particular case when the
subsystem $A$ denotes a single qubit (say the $j$th qubit) and $B$ denotes
the remainder of the system.
For this case we will write $\rho_j$ for $\rho_A$ and $\EE_{j}(\rho)$ for
$\EE_{AB}(\rho)$. 
In such an instance we have
$\EE_j(\rho)=-\lambda_j^{+} \log (\lambda_j^{+}) -\lambda_j^{-}
\log \lambda_j^{-}$ where $\lambda_j^{\pm}$ denote the two eigenvalues 
of $\rho_j$.
Using the fact that ${\rm tr} (\rho_j)=1$ and that the eigenvalues of
$\rho_j$ lie in the interval $[0,1]$ means that we can always
parameterise them as
$\lambda_j^{\pm}= \left(1\pm \sqrt{1-C^2_j}\right)/2$
with $C_j\in [0,\,1]$. 
For a two-qubit system $C=C_1=C_2$ is precisely the concurrence
\cite{hw},
so it is natural to call the $C_j$ for the $L$-qubit system the
local concurrences (cf. \cite{ckw}).
In terms of Pauli matrices it can be determined that 
$\rho_j=\left(I + \sum_{\alpha=x,y,z} \left<\sigma_j^\alpha\right> 
\sigma^\alpha_j\right)/2$ 
giving the local concurrence as 
\bea C_j=\sqrt{1-\sum_{\alpha=x,y,z} \left<\sigma_j^\alpha\right>^2}.\label{lc} \eea 

For the ground state of the BCS model we will establish a
correspondence between each local concurrence and a certain correlation
function $\C_j$ (see (\ref{cf}) below) describing the fluctuation in the
Cooper pair occupation numbers.  
An advantage of this approach is that it
applies equally to the thermodynamic limit where gauge symmetry is broken,
and for finite systems where there is no broken symmetry.  
We obtain
analytic results for the ALC in two extreme cases; (i) the thermodynamic
limit and (ii) the case of a single Cooper pair in a system with two
single particle energy levels.  
Between these
extremes we investigate the ALC in terms of the exact solution of the
model provided by the Bethe ansatz \cite{r}, which facilitates the
calculation of correlation functions \cite{zlmg,lzmg} and in turn the ALC.

{\it The reduced BCS model.}   
The reduced BCS Hamiltonian has received much attention as a
result of the effort to  
understand pairing correlations in 
nanoscale metallic systems \cite{ds,vd}. 
The Hamiltonian reads
\bea H_{\rm{BCS}}&=&\sum_{j=1}^{L}\e_jn_j
-d\lambda \sum_{j\neq k}^{L}
c_{j+}^{\d}c_{j-}^{\d}c_{k-} c_{k+}. \label{bcs}
\eea 
Above, $j=1,\dots,{L}$ labels a shell of doubly degenerate single
particle
energy levels with energies $\e_j$, $d$ is the mean level spacing and
$\lambda$ is the dimensionless coupling.
The operators $c_{j\pm},\,c^{\d}_{j\pm}$ are the annihilation
and creation operators for electrons at level $j$ where the labels $\pm$
refer to pairs of time-reversed states and $n_j=c^\dagger_{j+}c_{j+}
  + \ c^\dagger_{j-}c_{j-} $ is the
   electron number operator for
    level $j$.  
Throughout we will only consider 
systems at half-filling. 
The physical properties predicted 
by the model are quite
    different in the superconducting (SC) regime 
    ($d\ll\bg$, where $\bg$ is the bulk gap, defined below) 
    and the fluctuation
    dominated (FD) regime ($d\gg \bg$), the latter being the case for
    nanoscale systems. 
For the SC regime the variational BCS ansatz using
mean-field theory can be used to determine the ground state properties. 
However, the
mean-field approximation is not justified in the FD regime, 
where quantum fluctuations are significant.

The condensation
energy $E^c$ is defined as the energy loss relative to the energy of the
uncorrelated state (i.e. the ground state energy at $\lambda=0$, or
equivalently the energy expectation value of the Fermi sea).
In
the language of \cite{ddb}, it is equivalent to the entanglement gap for
this model. 
Intuitively, the entanglement gap gives an indication of
the ground state entanglement of the system. 
By definition, 
it is zero if and only if
the ground state is not entangled. 
It is thus desirable to determine
how the entanglement gap (or equivalently the condensation energy) 
relates to EMs.
It is known that the condensation energy is extensive in the SC regime, 
intensive in the FD regime, but with entirely smooth 
behaviour in the crossover \cite{ds,vd}. 
Below we will use a relation between condensation energy 
and the ALC to establish that a threshold coupling 
exists which marks 
qualitative differences in the ground state in terms of entanglement.  
An important 
quantity in our subsequent analysis will be played by the dimensionless
condensation energy per electron, defined by $\tE=E^c/\omega_D L$.

Next we discuss the decomposition of the Hilbert space into 
subsystems. 
At each energy level $\e_j$ there are four independent
states;
$\left|0\right>,\, c^\dagger_{+}\left|0\right>,
\,c^\dagger_{-}\left|0\right>,\, c^\dagger_{+}
c^\dagger_{-}\left|0\right>$.
These states serve as a \emph{ququadrit}, which can be further decomposed
into two qubits  
through the identification $\left|0\right>
\equiv\left|00\right>,\,
c^\dagger_{+}\left|0\right>\equiv\left|10\right>,
\,c^\dagger_{-}\left|0\right>\equiv\left|01\right>,\, c^\dagger_{+}
c^\dagger_{-}\left|0\right>\equiv\left|11\right>$ \cite{z,zw}. 
In the ground state all electrons are paired, so there is zero probability
of observing a single electron at any level $\e_j$. 
Thus for the ground state
each level  
$\e_j$ gives rise to a two-state system with basis 
$\left|0\right>,\,c^\dagger_{+}
c^\dagger_{-}\left|0\right>$, and each ququadrit serves as an effective qubit.

{\it The grand canonical ensemble.} 
The conventional BCS theory \cite{bcs} employs a grand canonical ensemble, 
where the electron number is not fixed, using the well-known variational ground
state ansatz 
\bea \left|\rm{BCS}\right>=\prod_{j=1}^L(u_jI+e^{i\theta}
v_j c_{j+}^{\d}c_{j-}^{\d}) \left|0\right> \label{var} \eea 
with $u_j,\,v_j$ real and satisfying $u_j^2+v_j^2=1$.
Including only those levels within the
cut-off given by the Debye frequency $\omega_D$ (i.e. $|\e_j|\leq \omega_D$
where the Fermi level is $\e_F=0$),
minimisation of the expectation value of the energy for  
(\ref{var}) gives 
\bea 4u_j^2v_j^2&=&{\D^2}/{(\e_j^2+\D^2)} \label{dist} \eea
where $\D=\omega_D/\sinh(1/\lambda)$ is the bulk gap. 
Unless stated otherwise 
we assume that the levels $\e_j$ are uniformly distributed. 
It can then be deduced that $\tE=(\coth(1/\lambda)-1)/2$.

With respect to the decomposition of (\ref{var}) into ququadrits  
discussed above, it is
a product state and thus not entangled.
There is however entanglement within each ququadrit 
subsystem. 
Expressing the state of a ququadrit as $u\left|00
\right> +v\,e^{i\theta} \left|11\right>$ it is easily determined that the
concurrence is $C=2|uv|$. 
Thus for the state (\ref{var}) we may define
$L$ local concurrences $C_j$ associated with each level $j$, {\it which quantify
the entire entanglement content of the state}. 
Remarkably, most of the $2^{L{+}1}{-}3L{-}2$ independent EMs are zero.
Note that the
definition of local concurrence here is not the same as the definition
of partial concurrence in \cite{md} 
based on the notion of broken time-reversal
symmetry of (\ref{var}) (see also \cite{g}).
A simple choice for an EM which reflects the overall ground state
entanglement is the ALC, 
$ \oC= {L^{-1}} \sum_{j=1}^L C_j$.
We remark that $\oC$ only quantifies bi-partite entanglement,
and not multi-partite entanglement \cite{ckw,lps,kem}.
However (\ref{var}) has no multi-partite entanglement. 
In the thermodynamic limit
$d\rightarrow 0$ with $Ld=2\omega_D$ finite 
we can compute the ALC as 
\bea \oC&=& \frac{\int_{-\omega_D}^{\omega_D} f(\e) \mu(\e) d\e 
}{\int_{-\omega_D}^{\omega_D}  \mu(\e) d\e} \label{alc} \eea   
with $f(\e)=\D/(\sqrt{\e^2+\D^2})$ and $\mu(\e)\geq 0$ a density
function for the distribution of the single particle energy levels.
For the case $\mu(\e)=1$ 
it is straightforward to determine that the 
ALC is
$\oC={1}/({\lambda\sinh(1/\lambda)})$.
Recalling \cite{vd} that the order parameter is given by 
$\Delta=\lambda d \sum_{j=1}^L \left<c_{j-}c_{j+}\right>$ 
we have 
$|\Delta|=\lambda \omega_D\oC$.  
This last relation reflects the fact that 
superconducting order arises from the instability of the Fermi sea 
due to Cooper pairing, which also results in the emergence of entanglement 
in the ground state.
Additionally we find $2{d \tE}/{d \lambda} = \oC^2$.   
For the canonical case (i.e. a finite system with fixed electron number)
the order parameter vanishes, but one can alternatively study the extent to
which `superconducting order' survives through the study of 
certain correlation functions \cite{vd}, which we will show can be used
to compute the ALC. 

\begin{figure}
\includegraphics[scale=.27]{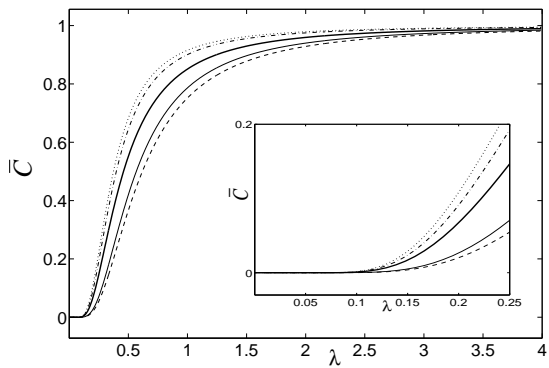}
\caption{\label{fig1} 
The ALC as a function of the dimensionless coupling $\lambda$
in the thermodynamic
limit for various distributions $\mu(\e)$ of the single particle levels.
The results shown are for, in increasing order of local density about the
Fermi level, $\mu(\e)=\e^2$ (dash),
$\mu(\e)= |\e|$ (solid), $\mu(\e)=1$ (bold),
$\mu(\e)=\omega^2_D-\e^2$ (dash-dot) and $\mu(\e)=\omega_D-|\e|$ (dot).
For each of these cases the
expression (\ref{alc}) can be evaluated analytically.
}
\end{figure}

We introduce the correlators describing the fluctuation in Cooper
pair occupation: 
\bea \C_j&=&\sqrt{\left< n_j^2\right>- \left<n_j\right>^2}
=\sqrt{\left<2-n_j\right>\left<n_j\right>}.\label{cf}
\eea
These correlators can be directly evaluated for the variational
wavefunction (\ref{var})  giving
$\C_j=2\left|u_j v_j\right|=C_j$, so the local concurrence is precisely
the fluctuation of the Cooper pair occupation.  
These 
fluctuations are localised within the range $\bg$ of the Fermi level
as given by (\ref{dist}). The behaviour of the ALC is strongly
influenced by the density of levels about the Fermi level, as depicted in 
Fig.~1. 
Hereafter we will only consider the case of uniform distribution 
of the levels to simplify the analysis.

In the
thermodynamic limit (\ref{var}) becomes the exact ground state and the 
finite size corrections are of order $1/L$ \cite{a}.
For systems with large but finite electron number in the SC regime,  
the entanglement of the state (\ref{var}) and the entanglement of the
ground state will be the same up to corrections of order $1/L$. 
However, the ground state cannot be accurately approximated by  
the product state (\ref{var}) in the FD regime. 
In this case the correlations 
spread out in energy space over the entire width $2\omega_D$
about the Fermi level \cite{vd}. 

{\it The canonical ensemble.} 
Our next step is to establish that the correlators (\ref{cf}) 
are still equivalent to
the local concurrences for a canonical system.
Recall that since there are no unpaired electrons in the ground
state, we can treat each ququadrit as an effective qubit. 
We express the action of the canonical Fermi algebra 
on the subspace of the Hilbert space with no unpaired
electrons in terms of the Pauli matrices  
through the identification 
$n=I+\sigma^z,~ c_{+}^{\d}c_{-}^{\d}=\sigma^+,~ 
c_{-}c_{+}=\sigma^-$.    
The uniqueness of the ground state and the $u(1)$ invariance of the  
Hamiltonian (\ref{bcs}) due to conservation of total electron number means 
that (\ref{lc}) reduces to 
$C_j=\sqrt{1-\left<\sigma^z\right>^2}$.  
Next we express the correlators (\ref{cf}) in terms of the Pauli
matrices, with the result being 
$\C_j= C_j$.    
A clarifying point is needed here. 
The local concurrence
$C_j$ is a measure of the entanglement between the effective qubit
associated with the $j$th level and the remainder of the system. 
Treating
the $j$th level as a ququadrit, the reduced density matrix becomes 
\bea \rho_j=\frac{1}{2}(2-\left<n_j\right>)\left|00\right>\left<00\right| 
+\frac{1}{2} 
\left<n_j\right> \left|11\right>\left<11\right|. \label{red} \eea 
Taking the partial trace over either qubit yields 
a reduced density matrix 
which has the same non-zero eigenvalues as (\ref{red}), 
so the local concurrence for either qubit within a ququadrit is the
{\it same} as the local concurrence of the ququadrit viewed as an
effective qubit. 

We see that the definition for the ALC
in terms of the correlators
(\ref{cf}) is the same for the canonical and grand canonical cases. 
Note that the derivation of the ALC in terms of (\ref{cf})   
for canonical systems relied on $u(1)$ invariance. 
In the
thermodynamic limit the $u(1)$ invariance of the ground state density
matrix is broken, but the same expression for the ALC  in terms of
(\ref{cf}) is valid.

\begin{figure} 
\includegraphics[scale=.27]{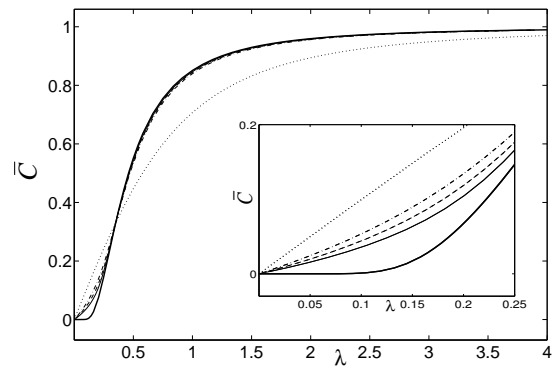}
\caption{\label{fig2}
 Ground state ALC for systems of $L=24$ (dot-dash), $40$ (dash) and $68$
(solid) levels. The
 bold and dotted lines show the analytic results for the
thermodynamical limit
 and $L=2$ respectively. 
The inset, showing the result for small $\lambda$, highlights that the ALC is 
$L$-dependent and not universal.}
\end{figure}

\begin{figure}
\includegraphics[scale=.27]{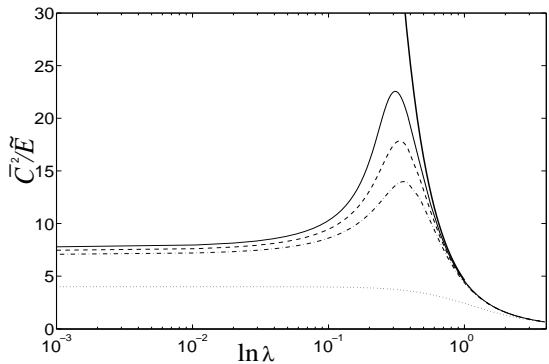}
\caption{\label{fig4}
The ratio $\oC^2/\tE$ versus $\ln (\lambda)$. The results shown are for
$L=24$ (dot-dash), $40$ (dash) and $68$
(solid) levels, while the bold and dotted curves 
are the analytic results obtained for
the thermodynamic limit and $L=2$. The maximum for each case is a 
threshold coupling, below which other EMs become significant. For $L=2$ and the thermodynamic limit
the maximum occurs at $\lambda=0$, as there is only bi-partite 
entanglement in these cases.  
}
\end{figure}

To analyse the ground state ALC in the canonical case we employ 
results from the exact solution \cite{r}.
An eigenstate of (\ref{bcs}) with $M$ Cooper pairs is characterised
by a set of complex parameters $\{v_1,\dots,v_M\}$ 
which provide a solution to the
set of Bethe Ansatz Equations (BAE) 
\bea 
\frac{2}{d\lambda}+ \sum_{k=1}^{L}\frac{1}{v_i-\e_k}
&=&\sum_{j\neq i}^M \frac{2}{v_i-v_j},
\label{bcsbae} \eea 
and the energy is given by $E=2\sum_{j=1}^M v_j+\lambda d M$. 
For the simple case of $L=2,\,M=1$,  
(\ref{bcsbae}) can be solved analytically yielding the ground state energy
$E= \e_1+\e_2-\sqrt{d^2\lambda^2+(\e_1-\e_2)^2}$. 
Using the Hellmann-Feynman theorem the correlators (\ref{cf}) can be
computed which gives the ALC as 
$\oC=\lambda/{\sqrt{1+\lambda^2}}$.

For the case of general finite $L$ the ALC can be computed through
determinant representations of the $\C_j$ which are 
given as functions
of the set $\{v_1,\dots,v_M\}$. 
The explicit formulae can be found in
\cite{zlmg,lzmg}. 
In the limits of weak \cite{f} and strong \cite{yba} 
coupling for large but finite 
$L$ we have the asymptotic
results 
\bea 
\tE&\sim&    \lambda/2(1+O(1/L)),\nn \\  
\oC&\sim&  1-1/(6\lambda^2)(1+O(1/ L)), 
~~~~~~~~~~~~~~~~~\, \lambda^{-1}\ll 1,   \nn \\ 
\tE&\sim&  \lambda^2\ln(2)/L(1+O(1/\ln L)),\,\,\nn \\
\oC&\sim& \lambda \sqrt{{2}/{L}} \,\ln (3+\sqrt{8})(1+O(1/\ln L)), 
\,\ \ \, ~~~~~\lambda \ll 1.  \nn \eea 
It is found that for
$\lambda^{-1}\ll1$, $\oC^2/\tE\sim 2/\lambda$,   
and 
~~\\
\bea \oC^2/\tE &\sim&   
(2\ln^2(3+\sqrt{8}))/\ln(2) 
\approx 8.97 
\nn \eea 
for $\lambda\ll 1$. 
Therefore in the FD and SC regimes the the quantity $\oC^2/\tE$ displays
scaling behaviour, i.e. the leading term is independent of $L$. 
In contrast to the FD and SC regimes,
which are characterised by the scales
$1/\ln (L)$ and $1/L$ respectively, the scale $1/\ln^2 (L)$ occurs for
the crossover regime~\cite{f}.

In Fig.~2 we show the ALC versus $\lambda$ for $L=2,\,24,\,40,\,68$
and the thermodynamic limit.
Like the condensation energy,
it is clear that the ALC is a smooth monotonic function of
$\lambda$  as
it crosses from the FD to the SC regime.
In Fig.~3 we plot the ratio $\oC^2/\tE$ versus $\ln (\lambda)$, 
for the finite cases
$L=24,40,68$. 
For sufficiently small 
$\lambda$ this ratio is approximately the constant value 
8.97 (up to a small correction $O(1/\ln (L))$) in agreement with the 
asymptotic result, but is nonetheless monotonically increasing. 
At sufficiently large $\lambda$ the ratio is monotonically decreasing 
and is well approximated by
the analytic curve for the thermodynamic limit. 
For $L=2$ and the thermodynamic limit the curves, also shown,
are monotonic, while in each of the other three cases 
there is clearly a maximum at a finite value of $\ln(\lambda)$. 
The coupling at which the maximum of $\oC^2/\tE$ 
occurs is a threshold: 
below this coupling  
(\ref{var}) no longer approximates the ground state and we must 
expect that other (generally multi-partite)  
EMs become significant.   

For larger values of $L$, we 
appeal to a heuristic argument based on the observation made 
in~\cite{f}: 
for the crossover regime, the condensation energy is roughly 
reproduced by simply summing up the contributions from
the perturbative 
result in the FD regime and the BCS mean field theory 
in the SC regime  \cite{gg}. This
also applies to the ALC.  Therefore,  
the same picture we have drawn from exact Bethe ansatz solutions
for small $L$ works for very large $L$, thus filling 
in the gap between the tractable but relatively small
$L$'s and the thermodynamic limit.  As $L$ becomes very large, 
the threshold coupling tends to
the value $2/\ln (L)$ (the coupling at which
mean-field theory breaks down~\cite{f}). This gives 
$\oC^2/\tE\sim  
\ln^2 (L)$ at the threshold, showing  
the peak in Fig.~3 is not bounded as $L$ increases.  
The competition of different scales in the crossover 
regime leads to the breakdown of 
the scaling behaviour  
observed in the FD and SC regimes.

We gratefully 
acknowledge financial support from the Australian Research
Council.


\begin{thebibliography}{99}
\bibitem{gg} Other fittings are available, see e.g. 
Eq. (24) in Ref. \cite{f}.
However, this does not affect our qualitative discussion.
\end{thebibliography}
\end{document}